\documentstyle[epsfig,twocolumn,aps,amstex,amssymb]{revtex}
\begin{document}
\draft

\title{Vortex Rings and Mutual Drag in Trapped Bose-Einstein Condensates}
\author{B. Jackson, J. F. McCann\cite{byline}, and C. S. Adams}
\address{Dept. of Physics, Rochester Building, University of Durham,
South Road, Durham, DH1 3LE, UK.}
\date{\today}
\maketitle

\begin{abstract}
We study the drag on an object moving through a trapped Bose-Einstein 
condensate, and show that 
finite compressibility leads to a mutual drag, which is subsequently 
suppressed by the formation of a vortex ring.
\end{abstract}


The appearance of vortices in quantum fluids has important consequences for 
many physical systems \cite{till,kibb76}. In particular,
there appears to be a direct link between vortex formation and the breakdown
of superfluidity in liquid helium \cite{donn}. However, a quantative 
comparision between theory and experiment is impeded by the complexity of this
many-body system. 

The recent observation of Bose-Einstein Condensation (BEC) in trapped alkali 
gases \cite{ande95}, on the other hand, has
yielded an ideal testing ground for many-body theories. Dilute condensates
can now be produced at temperatures far below the BEC transition, allowing an
accurate description by the Gross-Pitaevskii (GP) equation. By studying this
equation, insight can be gained into the relationship between vortex formation
and drag in superfluid systems. For homogeneous fluid flow past an obstacle,
numerical studies show that below a critical velocity the fluid exerts no net
force on the object \cite{fris92}, whereas above this velocity vortices are
nucleated, leading to a pressure imbalance which creates drag \cite{wini98}.

The picture is less clear, however, in trapped Bose gases, which are
inhomogeneous and of finite size. Simulations have shown that dragging an
object, created by a far-detuned laser beam, through an inhomogeneous
condensate produces compression waves and a series of vortex pairs
\cite{jack981}. The formation of these structures will influence the 
dynamics of the condensates and the drag on the object. An attractive system
for studying drag is the multi-component condensate \cite{myat97}, where the 
magnetic sensitivity of the different hyperfine states allows the fluid
components to be intermingled or spatially separated. In this paper, we 
simulate  actively dragging a small condensate, the `object', through a 
larger, less tightly confined condensate, the `fluid'. We find that
a mutual drag arises due to deformation of the compressible object, which is 
relieved by the formation of vortex rings in the fluid. The time 
dependence of the mutual drag provides a signature of ring formation. This 
process is analogous to ring nucleation by moving ions in superfluid 
$^4{\rm He}$ \cite{rayf64}, albeit on a different scale. 

The numerical methods follow closely those employed in our previous work
\cite{jack981,jack982}. The coupled 
GP equations for the condensate
wavefunctions, $\psi_i (\mbox{\boldmath $r$},t)$, in the $x-y$ plane can be 
written as:
\begin{equation}
 i \partial_t \psi_i = \left[ -\nabla^{2} + V_i + C  
 \left( \alpha_i |\psi_i|^2 + |\psi_{j}|^2 \right) 
                                     - \mu_i \right] \psi_i,
\label{eq:GrossP}
\end{equation}
where $i,j=1,2$ $(i \neq j)$, $\mu_i$ are the chemical potentials at
equilibrium, $V_i$ are the trap potentials, $\alpha_i$ are the ratios of the
scattering lengths, and $C$ is the nonlinear coefficient \cite{num}.
We have performed 2D, cylindrically-symmetric 3D, 
and full Cartesian 3D simulations \cite{meth}.

To demonstrate that coupled GP equations give an accurate description of the
formation of structure in the condensate, we begin by modelling an experiment
on component separation \cite{hall98}. We assume that a condensate
initially in state $|1 \rangle$ with trapping potential
$V_1=(x^2+ \epsilon^2 y^2)/4$, is subject to an interaction which transfers 
$50\%$ of the population into state $|2 \rangle$, which experiences a 
displaced
potential $V_2 = (x^2 + \epsilon^2 (y-y_0)^2)/4$, where $y_0$ is the offset
\cite{para}. The displaced potential causes the two components to separate
and subsequently oscillate at large amplitudes. Nonlinear mixing \cite{rupr95}
leads to excitation of high-order modes, which appear as structure
in the density profiles \cite{hall98}. Fig.\ 1(a) displays a
density cross-section at $t=8.52$ (corresponding to $65\,{\rm ms}$).
However, in the experiment, most of the structure had disappeared by this 
time. This disparity may be due to Landau damping at finite temperatures, where
condensate
excitations are absorbed by the thermal cloud. To simulate this
effect, we propagate the GP equations (\ref{eq:GrossP}) in complex 
time $t \rightarrow (1+i\Lambda) \tilde{t}$ \cite{choi98}.
Components with an energy, $E$, decay exponentially at a rate proportional
to the damping constant, $\Lambda<0$ \cite{damp}, and to
$E-\mu_i$, in agreement with theory for the collisionless 
low-energy regime \cite{liu97}. Higher frequency 
excitations are preferentially damped, reducing the density variation within 
the condensate (Fig.\ 1(b)).

Fig.\ 2 shows the center-of-mass position of the two condensates 
($Y_{i} = \iint y|\psi_i|^2\,{\rm d}x\,{\rm d}y$) as a function of time. 
Population of high-energy excitations results in damping of the dipole mode, 
in agreement with previous work \cite{sina98}. However, this mechanism alone 
is insufficient 
to explain the observed experimental damping. Propagation in
complex time results in further damping, but at a relatively slow rate.
We also investigate the effect of an offset in the $x$-direction, which may 
arise as a consequence of the experimental geometry \cite{cornpc}. In this 
case,
mixing between the fast ($y$) and slow ($x$) vibrational modes \cite{mix} could
be more significant than finite temperature effects in explaining the observed 
damping (see Fig.\ 2), though it is likely that a combination of these 
mechanisms is responsible.
\begin{figure}
\centering\epsfig{file=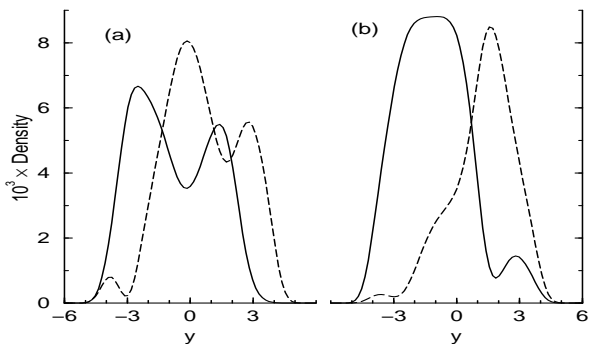, clip=, width=4.5cm, height=8.0cm,
 bbllx=100, bblly=60, bburx=585, bbury=645, angle=270}
\caption{Cross-sections through 2D density profiles, $|\psi_1|^2$ (dashed) and
 $|\psi_2|^2$ (solid), at $x=0$ and $t=8.52$, showing results of (a) undamped, 
 and (b) damped ($\Lambda=-0.025$) propagation.}
\end{figure}
\begin{figure}
\centering\epsfig{file=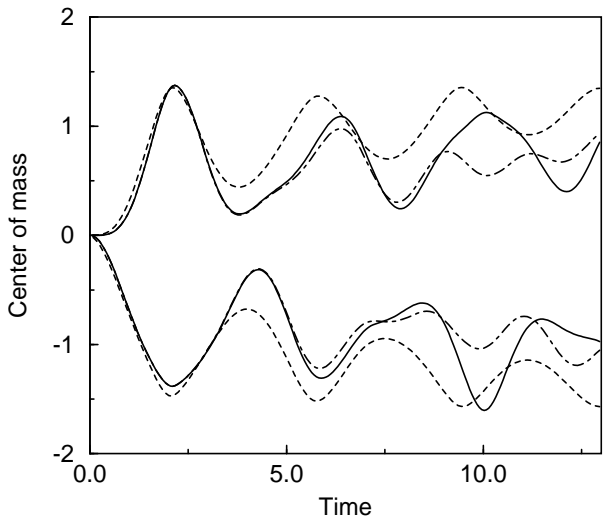, clip=, width=7.0cm, height=8.0cm,
 bbllx=90, bblly=45, bburx=575, bbury=610, angle=270}
\caption{Center-of-mass oscillations, $Y_{i} (t)$, in $|1 \rangle$ (top) and 
 $|2 \rangle$, for undamped (solid) and damped $\Lambda=-0.025$ (dashed)
 propagation. The dot-dashed lines show the effect on undamped propagation of 
 an additional offset, $x_0=0.240$, where decay of the oscillations are more 
 marked than even a large Landau damping.}
\end{figure}

As is apparent in Fig.\ 2, over short timescales ($t<2$) thermal damping is 
negligible and the undamped GP equations provide a reliable model.
We now turn to
the situation where one condensate flows through the other, in direct analogy 
with an object moving through a fluid. This may
be realized experimentally by using a magnetic trap to
confine atoms in state $|2 \rangle$, whilst an optical dipole trap, moving
with relative velocity $-v$, loosely confines atoms in a 
magnetically insensitive level $|1 \rangle$. We employ the coupled equations 
(\ref{eq:GrossP}), with $V_1=(x^2+y^2)/4$ and $V_2=V_1+(x^2+(y-vt)^2)$, and 
$\alpha_i=1$. The initial state is found using imaginary time
propagation \cite{esry97}. The repulsive mean field
arising from
the `object' ($|2 \rangle$) creates a local minimum in the center of the
density profile of the `fluid' ($|1 \rangle$) \cite{ho96}. 
The depth of the minimum depends on the interaction strength $C$, and on the 
fraction of atoms in the fluid, $f$.
\begin{figure}
\centering\epsfig{file=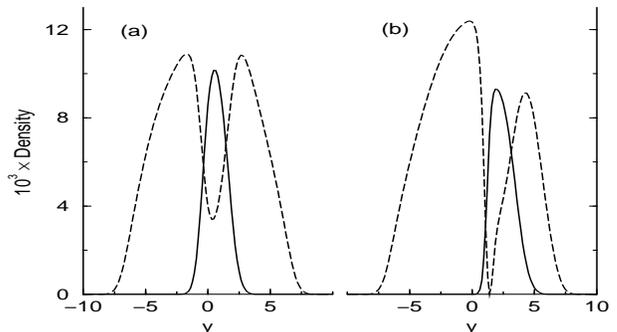, clip=, width=4.5cm, height=8.0cm,
 bbllx=90, bblly=45, bburx=585, bbury=635, angle=270}
\caption{Cross-sections through 2D density profiles at $x=0$. Condensate 
 $|2 \rangle$ (the `object',
 solid line) moves through $|1 \rangle$ (the `background fluid', dashed), 
 due to displacement of trap potentials at constant velocity, $v=3$; $C=1050$,
 $f=0.95$ (where $f$ is the fraction of atoms in $|1 \rangle$). Two 
 time-frames are shown: (a) $t=0.75$, where drag arises from a
 process analogous to phonon emission from the accelerating object, 
 (b) $t=1.3$, where deformation of the object and surrounding fluid leads to
 an additional drag.}   
\end{figure}

Displacement of the object potential ($V_2$) at $t>0$ induces motion of the
object, which leads to the minimum in the background fluid becoming 
progressively deeper at a rate which
increases with $v$ (see Fig.\ 3). When the density minimum reaches zero, it 
evolves into
a vortex ring. In addition, motion of the object creates a response in
the fluid, implying a back-action, or drag, on the object. The drag may be 
studied through the center-of-mass acceleration, given by:

\begin{equation}
 \ddot{\mbox{\boldmath $R$}}_i(t)=-2\int\limits_V |\psi_i 
 (\mbox{\boldmath $r$},t)|^2
 \nabla\mu_i(\mbox{\boldmath $r$},t)\,{\rm d}^3\mbox{\boldmath $r$}.
\label{eq:Acc}
\end{equation}
For steady flow of an homogeneous condensate without vorticity, the integral 
vanishes, implying superfluidity \cite{nozi90}. However, for a finite 
inhomogeneous system there is no steady state and a different criterion for 
superfluidity should be sought.
There are four energy terms that contribute to the acceleration; however, the 
kinetic and self-interaction terms are found to be negligible, leaving only:

\begin{equation}
 \ddot{Y}_{i,{\rm trap}}(t)=-2 \iint |\psi_i|^2 \frac{\partial V_i}
 {\partial y}\,{\rm d}x\,{\rm d}y,
\label{eq:TrapAcc}
\end{equation}
and,
\begin{equation}
 \ddot{Y}_{i,{\rm mut}}(t)=-2C \iint |\psi_i|^2 \frac{\partial}
 {\partial y} |\psi_{j}|^2\,{\rm d}x\,{\rm d}y,
\label{eq:MutAcc}
\end{equation}
which correspond to the contributions from the trap potentials and the 
interaction between condensates, respectively.

The mutual acceleration (\ref{eq:MutAcc}) is particularly interesting in
addressing the issue of drag.
Fig.\ 4 shows the calculated mutual acceleration as a function of time.
At later times ($t>2$) the condensates repel due to the inhomogeneity in
the background fluid. For $t<2$ the acceleration is negative,
implying an effective attraction, or equivalently a drag on the object. 
\begin{figure}
\epsfig{file=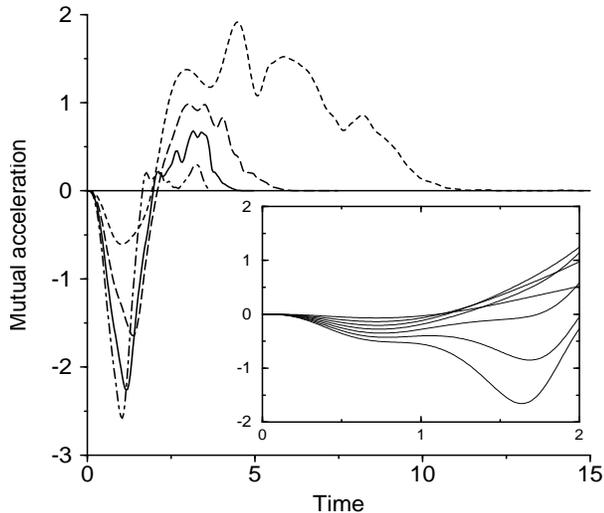, clip=,width=7.0cm, height=8.0cm,
 bbllx=65, bblly=15, bburx=600, bbury=600, angle=270 }
\caption{Plot of $\ddot{Y}_{2,{\rm mut}} (t)$ for $C=600$, $f=0.83$;
 $v=1,2,3,4$. The `drag' is evaluated from the minima of the curves, which 
 increases in magnitude as the velocity rises. Inset shows a similar plot 
 ($C=1050$, $f=0.95$) for $v=0.25$ to $v=1.75$ in steps of $0.25$. 
 At low $v$ the local minima arise due to the finite response of the background
 fluid, $|1 \rangle$. At higher $v$ an additional minimum appears at $t>1$ due 
 to the compressibility of the object. The position of each minimum 
 coincides with the moment of vortex formation.} 
\end{figure}

For $t<1$ and low velocities, Fig.\ 4 (inset),
the force results from the slow response of the fluid to the object
acceleration. The object moves to the front of the potential well it  
creates in the mean field of the fluid, resulting in a restoring
force which persists until the fluid can respond (Fig.\ 3(a)). The maximum 
attractive drag
(i.e.\ the minima in Fig.\ 4) is plotted as a function of velocity in Fig.\ 5.
This process, which is equivalent to phonon emission by an accelerating 
object, is responsible for the linear section of the drag curve.   

For velocities near the speed of sound in the object,
$c_{2}=\sqrt{2C|\psi_2|^2}$, the object begins to deform, with the result that
the overlap between the
two fluids is reduced behind and enhanced in front (Fig.\ 3(b)). This 
increases the drag, producing 
the additional minima at $t>1$ in Fig.\ 4 (inset). Onset of this process is 
rapid with increasing velocity, giving rise
to a sharp transition in the drag curve. This transition can be compared to 
that between superflow and normal flow in the homogeneous system 
\cite{fris92,wini98}. However, due to the finite size of the background fluid 
there is no steady flow condition, and drag is produced even at low 
velocities. 
\begin{figure}
\centering\epsfig{file=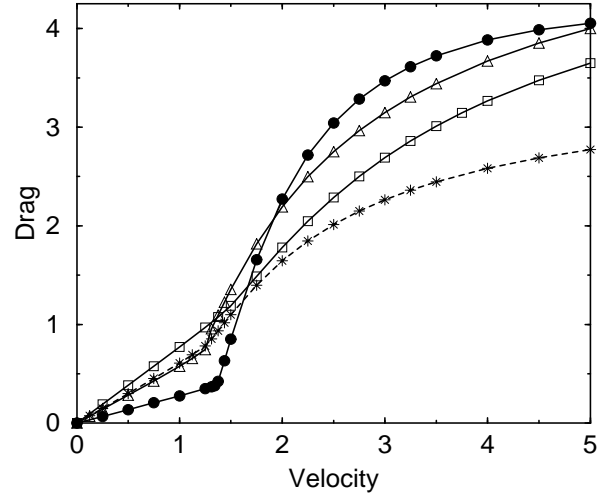, clip=,width=7.0cm, height=8.0cm, 
 bbllx=95, bblly=60, bburx=600, bbury=630, angle=270 }
\caption{Peak mutual attraction (drag) as a function of velocity. Curves are
 shown for: $\ast$ $C=600$, $f=0.83$;  $\vartriangle$ $C=1100$, $f=0.91$;
 $\square$ $C=1200$, $f=0.83$; $\bullet$ $C=1050$, $f=0.95$. The curves show 
 linear dependence of drag at low velocities, and enhanced drag at high 
 velocity due to the compressibility of the condensates. The magnitude of the 
 additional drag is enhanced for highly compressible object condensates 
 i.e.\ when $1-f$ is small, or when the mean-field interaction 
 between condensates is large.}
\end{figure} 

The drag force increases with time until the local minimum in the background 
fluid reaches zero (Fig.\ 3(b)), from which a vortex ring is formed. 
Subsequent expansion of the ring results in the condensate minimum 
being filled, thereby decreasing the pressure imbalance across the object. The 
object
returns to a more symmetric shape, and the drag decays, as is
apparent in Fig.\ 4. Hence, vortex formation in this context tends to limit or 
reduce the drag. 

In summary, we simulate a scheme where one condensate is
pulled through another, and identify mutual drag effects, and vortex ring 
creation which acts to suppress drag. Although vortex rings would be
difficult to detect directly, we have shown that they produce a significant
change in the mutual drag which could be observable.

Financial support for this work was provided by EPSRC. We are grateful to
E. A. Cornell (JILA) for providing us with information on the experiment
reported in \cite{hall98}.


\begin{references}
\bibitem{num}
 Throughout this paper, harmonic oscillator units (h.o.u.) are used, where for
 atoms of mass $m$ trapped in a harmonic trap of frequency $\omega$, the units
 of length, time and energy are $(\hbar/2m\omega)^{1/2}$, $\omega^{-1}$ and 
 $\hbar\omega$ respectively. As a result of this scaling, the wavefunction is 
 redefined so that the total density, $|\psi_1|^2+|\psi_2|^2$, is normalised 
 to unity, and $C_{2D}=8 \pi N a_{12}$, where $N$ is the
 total number of atoms per unit length along the $z$-axis. We define 
 $\alpha_1=a_1/a_{12}$ and $\alpha_2=a_2/a_{12}$, where the scattering
 lengths between atoms in the same hyperfine level are denoted by $a_1$ and 
 $a_2$, while $a_{12}$ represents scattering between hyperfine states.
\bibitem{meth}
 We used a Fast Fourier Transform 
 method, where a 2D wavefunction is typically discretized on a 
 $256 \times 256$ grid inside a $40 \times 40$ box, while for 3D we employed
 a $128 \times 128 \times 64$ grid. The time-step, $\Delta t$, 
 must be small enough to ensure stability over sufficiently long propagation  
 times; $\Delta t=5 \times 10^{-4}$ is adequate for most purposes.
\bibitem{para}
 Parameters in our simulations are chosen to reflect closely those 
 in the experiment \cite{hall98}. In particular, $\epsilon=\sqrt{8}$,
 $\omega=2\pi\times 59 /\sqrt{8}$, $C=3160$ and
 $y_0 = 0.240\,{\rm h.o.u.} \simeq 0.4\,\mu {\rm m}$. Ratios of the scattering
 lengths are taken to be $\alpha_1=1.03$ and $\alpha_2=0.97$, where
 $a_{12}=5.5\,{\rm nm}$.
\bibitem{mix}
 Due to the anisotropy of the trap,
 the dipole mode corresponding to oscillations along the `slow' $x$-axis
 possesses a lower energy than along $y$. Displacement of the trap initially 
 leads 
 to oscillations along the `fast' axis as previously; however, the broken 
 symmetry now results in transitions that gradually populate the lower level.
\end{references}
\end{document}